\documentclass[12pt,conference,onecolumn]{IEEEtran}
\usepackage{blindtext}
\usepackage{times}
\usepackage[pdftex]{graphicx}
\usepackage{amsmath,amssymb,amsthm}
\usepackage[caption=false,font=footnotesize,labelformat=simple]{subfig}
\usepackage{textcomp}
\usepackage[update,prepend]{epstopdf}
\usepackage{multibib}
\usepackage{url}
\usepackage{color}
\usepackage{multirow}
\usepackage{multicol}
\usepackage{cleveref}
\usepackage[bottom]{footmisc}
\usepackage[noadjust]{cite}

\usepackage{etoolbox}
\makeatletter
\patchcmd{\@makecaption}
{\scshape}
{}
{}
{}
\makeatletter
\patchcmd{\@makecaption}
{\\}
{.\ }
{}
{}
\makeatother

\makeatletter

\begin{document}
%
\title{Meeting Real-Time Constraint of Spectrum Management in TV Black-Space Access}
%
%
%

\author{Zhongyuan~Zhao
	~\IEEEmembership{Student Member,~IEEE,}\\
	Cyber-Physical Networking Lab\\Department of Computer Science and Engineering\\University of Nebraska-Lincoln, Lincoln, NE, 68588 USA\\e-mail: zhzhao@cse.unl.edu.
}

%



\maketitle


\IEEEpeerreviewmaketitle
\begin{abstract}
The TV set feedback feature standardized in the next generation TV system, ATSC 3.0, would enable opportunistic access of active TV channels in future Cognitive Radio Networks. This new dynamic spectrum access approach is named as black-space access, as it is complementary of current TV white space, which stands for inactive TV channels. TV black-space access can significantly increase the available spectrum of Cognitive Radio Networks in populated urban markets, where spectrum shortage is most severe while TV whitespace is very limited. However, to enable TV black-space access, secondary user has to evacuate a TV channel in a timely manner when TV user comes in. Such strict real-time constraint is an unique challenge of spectrum management infrastructure of Cognitive Radio Networks. In this paper, the real-time performance of spectrum management with regard to the degree of centralization of infrastructure is modeled and tested. Based on collected empirical network latency and database response time, we analyze the average evacuation time under four structures of spectrum management infrastructure: fully distribution, city-wide centralization, national-wide centralization, and semi-national centralization. The results show that national wide centralization may not meet the real-time requirement, while semi-national centralization that use multiple co-located independent spectrum manager can achieve real-time performance while keep most of the operational advantage of fully centralized structure.
\end{abstract}

\section{Background}
To address the spectrum crisis caused by rapidly growing wireless devices \cite{Akyildiz2006}, the Federal Communications Committee (FCC) allows unlicensed Television Band Devices (TVBDs) to operate in TV white spaces (TVWS) as secondary users (SUs) in the U.S. \cite{2ndOrder}. TVWS is defined as geographical areas where over-the-air TV services are unavailable \cite{2ndOrder}. Under TVWS paradigm, TVBD can access a list of TVWS channels obtained by querying a geolocation database. The geolocation databases calculate TVWS channels by approaches in \cite{5457914} based on the registration information of TV stations and radio propagation models. Since TV stations do not change their status often, geolocation databases only need to be updated on a daily basis, and a secondary user could also query only a few times per day if not roaming. Therefore, workload of a geolocation database can be handled by technology for large website. Actually, such services are provided by Internet companies, such as Google, Microsoft, and Spectrum Bridge. 

\begin{figure} [h!]
	\centering 
	\subfloat[]{
		\includegraphics[height=2in]{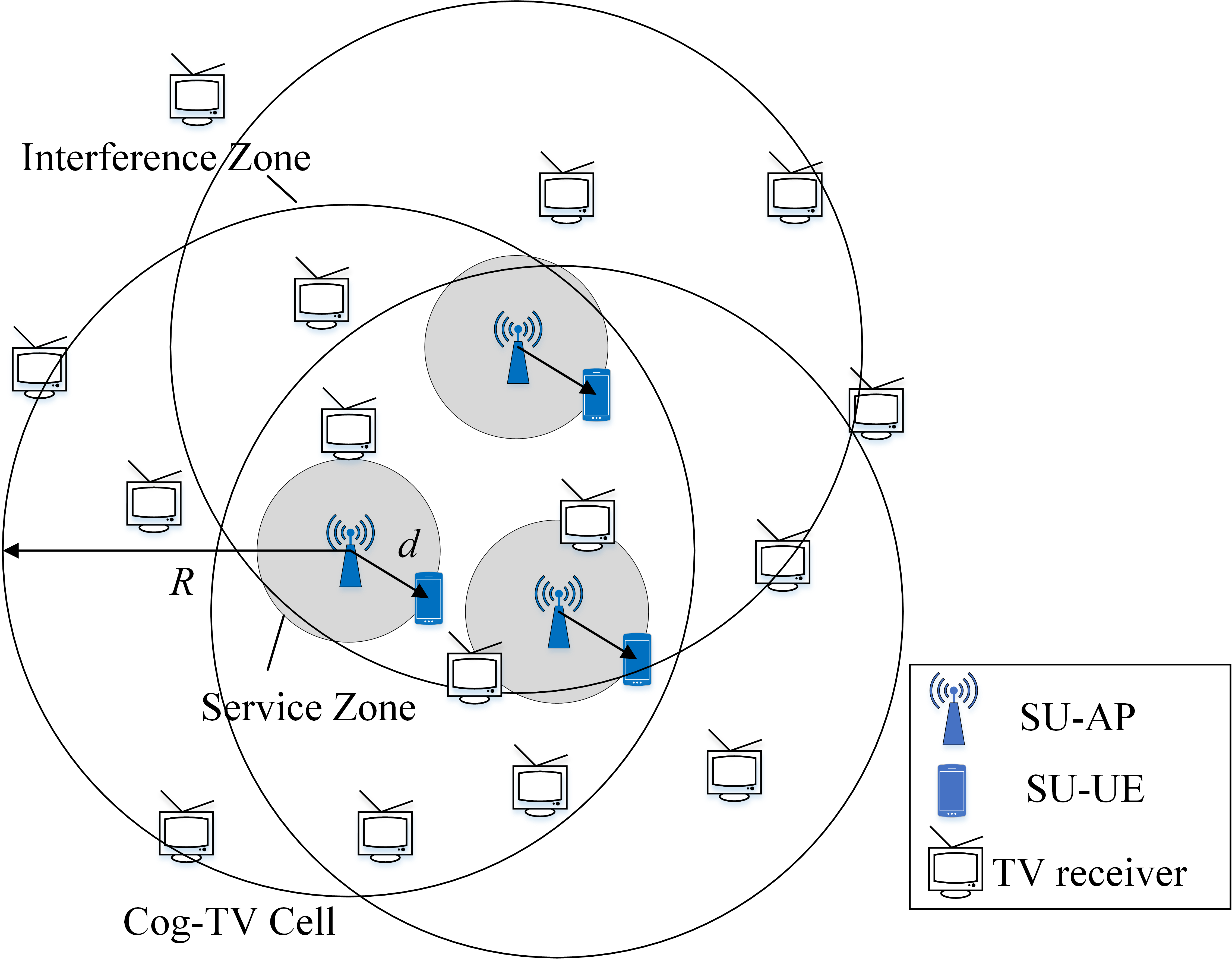}
		\label{fig:cell}
	}
	\subfloat[]{
		\includegraphics[height=1.2in]{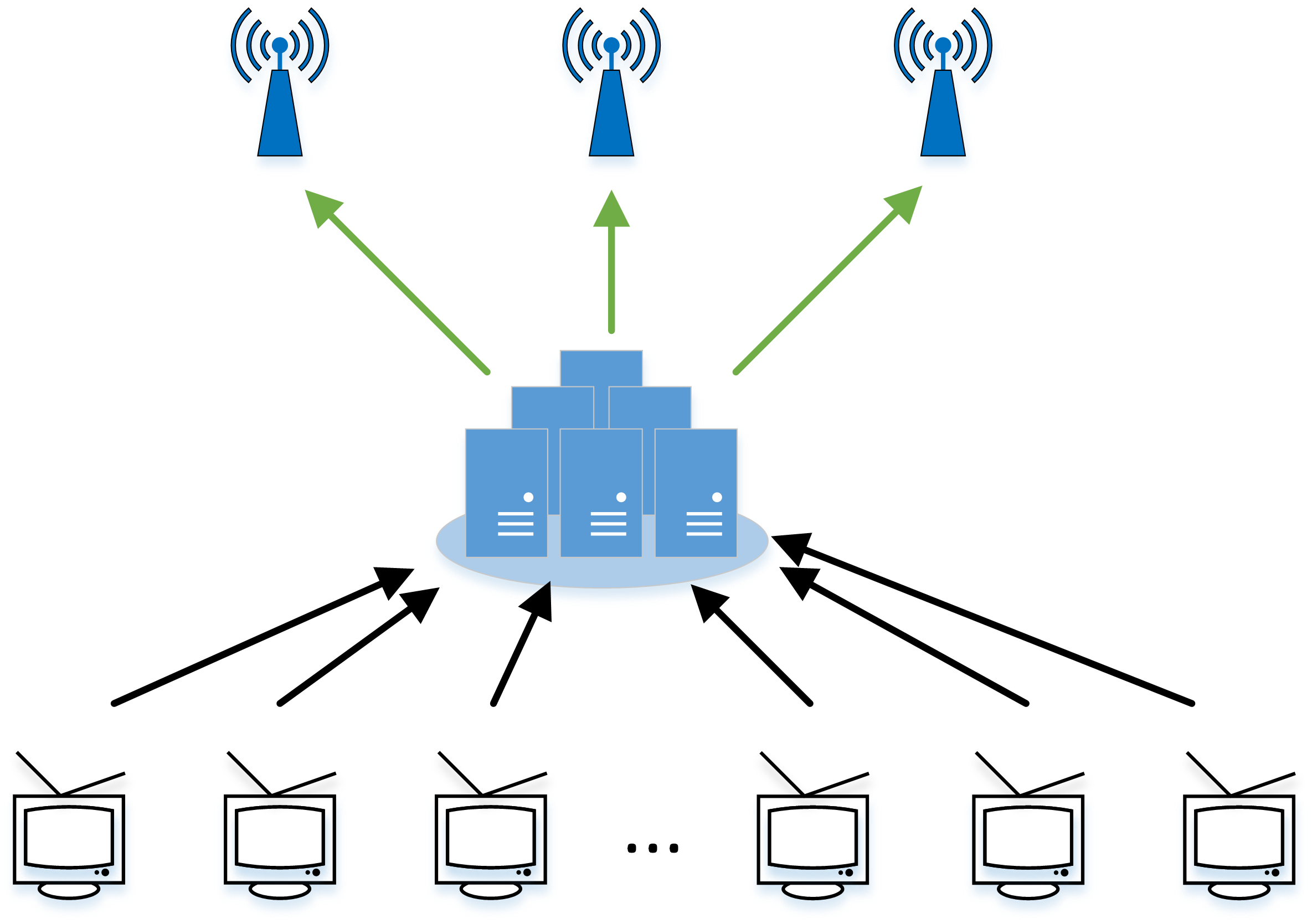}
		\label{fig:centralized}
	}
	\subfloat[]{
		\includegraphics[height=1.2in]{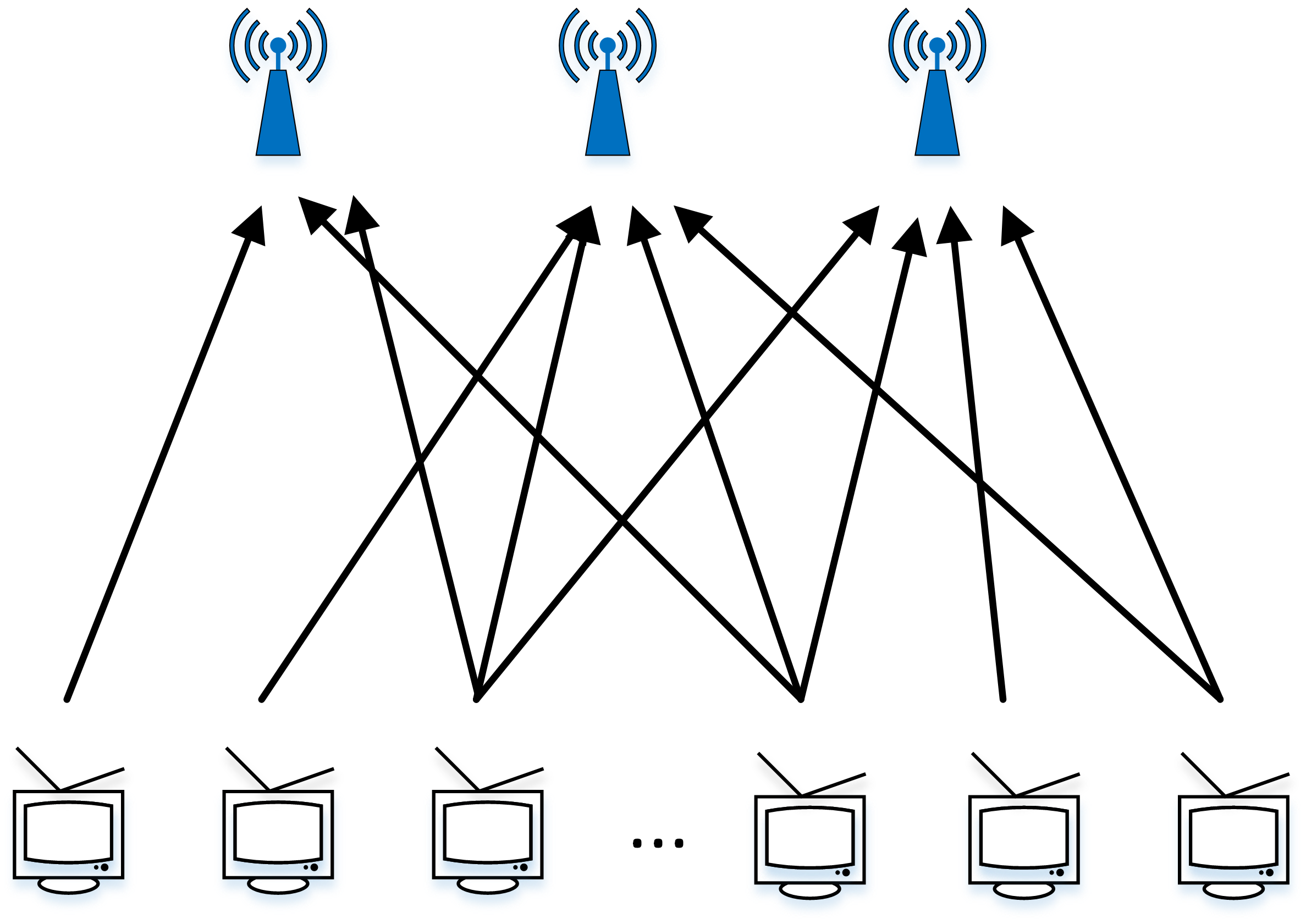}
		\label{fig:distributed}
	}
	\caption{(a) Cognitive Television (Cog-TV) Cellular Network, (b) Fully Centralized SM (c) Fully Distributed SM.}
	\vspace{-0.2in}
\end{figure}

However, the TVWS approach leaves populated urban areas, where spectrum shortage is most severe, with too few TVWS \cite{5457914,Hessar2014,zhzhao_techreport}, for these areas are also major markets of broadcast TV. As TV receiver feedback becomes essential in the next generation terrestrial TV standard, ATSC 3.0 \cite{Fay2016}, a TV receiver-facilitated TV spectrum sharing paradigm is proposed to exploit the underutilized legacy Over-The-Air TV service areas (TV black-spaces) \cite{Zhao14Globecom,Rempe2017,zhao2017}. 
An exemplary network structure of TV black-spaces paradigm is illustrated in Fig.~\ref{fig:cell}, where a secondary user base-station can schedule wireless communication on a list of free TV channels that are not watched by any TV receivers (primary users) in its interference zone.  This free channel list is dynamically determined by the behaviors of those TV receivers. The secondary user base-station has to evacuate communication on a TV channel immediately if any TV receiver in its interference zone is tuned in. 
Analyses based on empirical data of population, TV stations, TV usage and traffic patterns, show that the TV black-spaces approach can bring several folds increase of available spectrum in major urban areas compared to TVWS approach, meanwhile requires much stricter timing on spectrum management to protect the quality of experience of TV viewers  \cite{zhao2017}. It is a challenge to design infrastructures for the timing requirement of spectrum management in the TV black-spaces.

\section{Spectrum Management Infrastructure}
The spectrum management in TV black-spaces includes: receiving feedback from TV receivers when they change channels, update free channel list for a secondary base-station, coordinating neighboring secondary base-stations for spectrum reuse and mobility management. Because of the random individual choices of each TV viewer, the free channel list for each secondary base-station would be different and can be changed at any time. Delay is introduced by network and processing on the chains of TV set feedback, spectrum management, and resource scheduling at local secondary base-station. The real-time constraint is that secondary user must stop transmitting within 0.3 sec once a TV user push the button on the remote. 

The optimal solution could be anywhere between fully centralization (Fig.~\ref{fig:centralized}) and fully distribution (Fig.~\ref{fig:distributed}). The benefits of centralization include simplified TV receiver feedback, seamless across-cell coordination, low backhaul bandwidth overhead, and all the typical benefits of cloud computing. Fully distributed solution can reduce the latency for every chain of spectrum management, but would require more backhaul bandwidth consumption for TV set feedback and cross-cell coordination, and higher processing capability at the secondary base-station (cost). 

\section{Cyber Physical Systems-Based Approach}
In this project, we aim to find an approach to strike a balance between two extremes through Cyber-Physical Systems (CPS).  Parameters of the system include the dimensions of Cog-TV cell, and the characteristics of densities and traffics of TV receivers and secondary users, and network latency, and processing capacity of a spectrum manager processor and/or cluster (unit cost). The output is the number of secondary cells to be managed by a spectrum manager unit, and estimated performance and cost profile.

Based on approach of scheduling in CPS \cite{LiQiao10}, the physical network is modeled as a graph, where each node represent a secondary base-station. A spectrum manager unit is represented by a clique that connecting multiple nodes. The number of nodes in a clique is determined by the processing capacity of the spectrum manager unit, and the processing jobs at spectrum manager unit is modeled as a queueing model. The processing delays is then calculated based on the constraints of processing capacity and cost of the spectrum manager unit. The maximal size of each clique is subjected to the processing delays and network latency, and is constrained by the maximal allowable channel evacuation latency (e.g. 0.3 sec). 

Based on the graph and queueing model, metrics of performance (for both TV users and secondary users), and cost of secondary base-station and spectrum manager unit can be calculated as functions of size of clique. Numerical searching or analytical optimization algorithm can be used to find out the optimal size of clique. The results could be a function of traffic load, and thus the degree of centralization could be variable. This dynamic degree of centralization can reduce energy consumption by adaptive the spectrum manager control system to the traffic.

\section{System Model}\label{sec:cps}
We describe the system with a physical model and a cyber model. The physical model includes the events of primary users switching channel (called zapping event), the reactive action of channel evacuation by secondary users, and the latencies in the network and spectrum manager(s) between these two events. The cyber model describes the computational process within a spectrum manager, of which the delay depends on scale of the spectrum manager.
\subsection{Physical Model}\label{sec:cps:phy}
\begin{figure} [h!]
	\centering 
	\vspace{-0.2in}
	\subfloat[]{
		\includegraphics[width=0.45\linewidth]{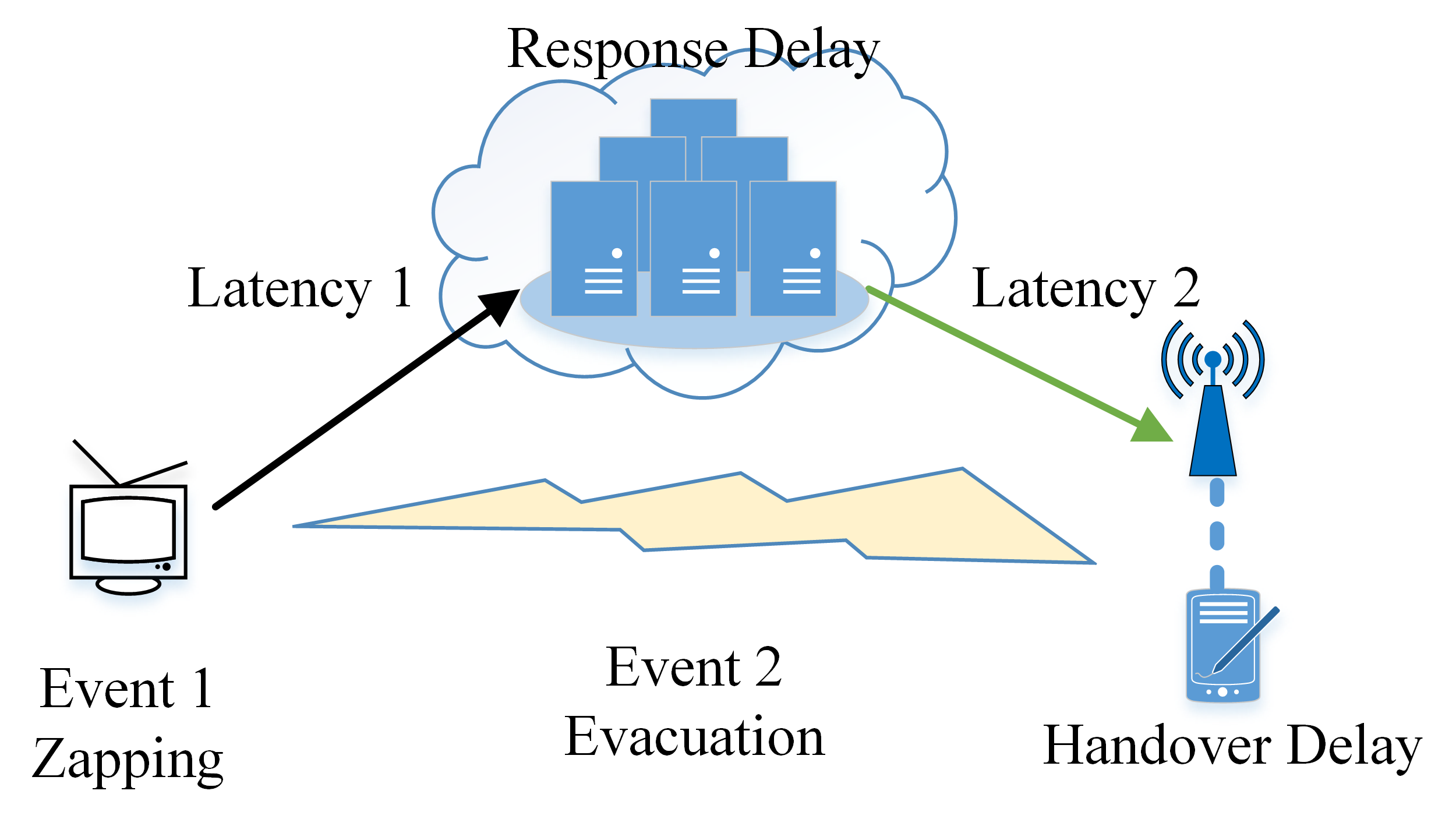}
		\label{fig:delay:sys}
	}
	\subfloat[]{
		\includegraphics[width=0.45\linewidth]{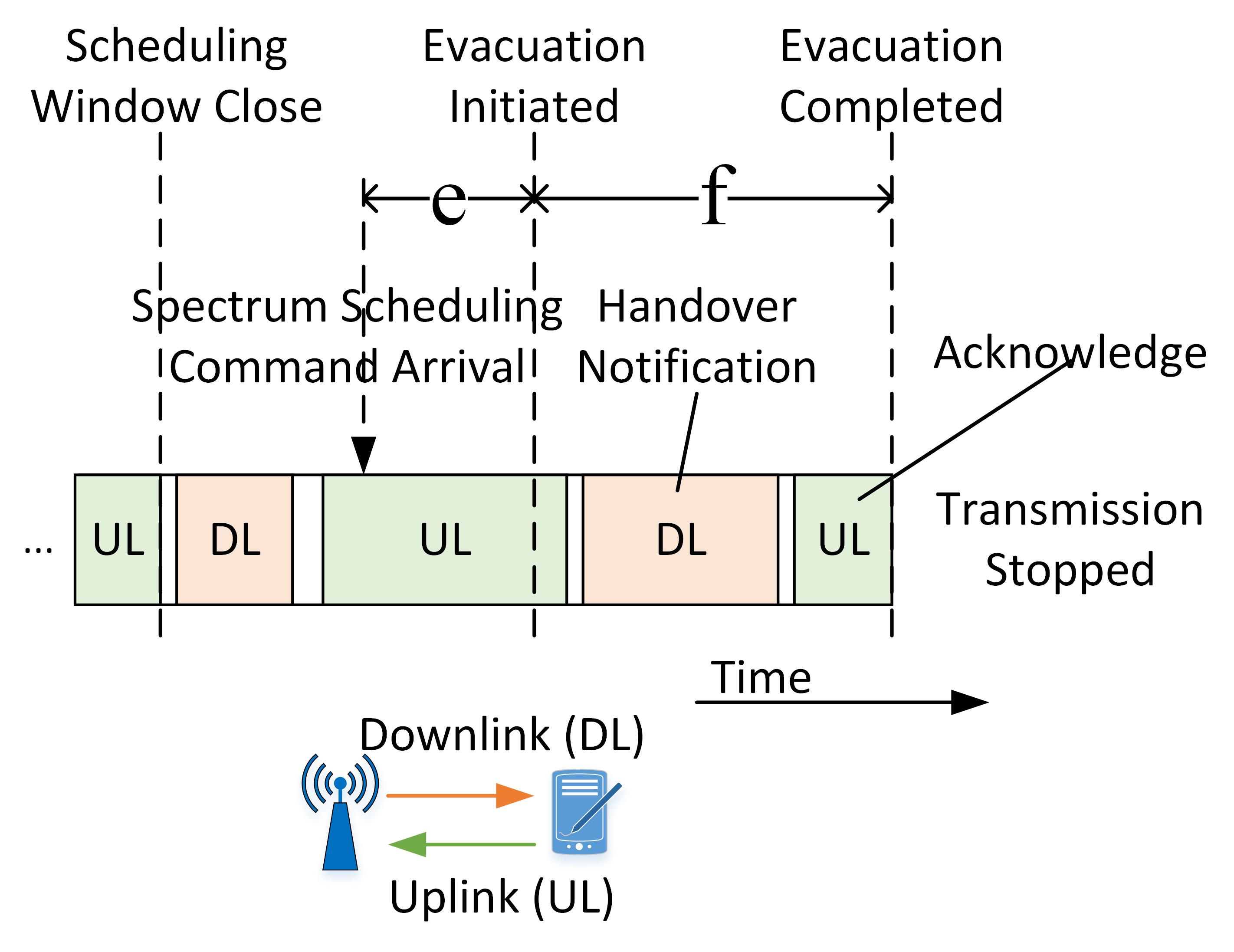}
		\label{fig:delay:su}
	}
	\vspace{-0.1in}
	\caption{(a) Physical Model, (b) Secondary User Delay.}\label{fig:delay}
\end{figure}
We first describe the physical process from the perspective of events, as illustrated in Fig.~\ref{fig:delay:sys}. The first type of event (Event 1) is zapping, which is a TV user initializing a switch of TV channel by pushing a button at his/her remote or TV sets. Event 1 triggers a signaling message being immediately sent to the spectrum manager by the TV set. The message contains the identity of TV set (which indicates its location) and the requested channel, $b$. The message of event 1 is sent to a spectrum manager, which checks if any secondary links in the vicinity of the requesting TV set are currently operating on the requested TV channel $b$. If yes, then the spectrum manager will immediately send a spectrum scheduling command to these secondary links. The spectrum scheduling command contains free channels that the interrupted secondary links could continually operate on. Once the command reaches target secondary link(s), the secondary base-stations (SBS) or base-station(s) will schedule attached secondary user equipments (SUEs) to switch from channel $b$ to a free channel provided by the spectrum manager. After these steps, the second type of event (event 2) happens, that is the secondary users evacuating the channel $b$. Once secondary links on channel $b$ are evacuated, the TV receiver can receive clean TV signal on channel $b$, and start to buffer and decode the TV signals.

From an event 1 to its reactive event 2, there are three types of delays: network latency, response delay of spectrum manager, and the reaction time of secondary link (handover delay), as shown in Fig.~\ref{fig:delay:sys}. Network latency includes the time a signaling message travels from TV user to the spectrum manager, as well as a spectrum scheduling command travels from the spectrum manager to the SBSs in general. Network latency depends on the physical distance between the TV user and spectrum manager, and follows a linear and stochastic model \cite{goonatilake2012}:
\begin{equation}\label{eq:network}
t_N(x) = ax+b + c\;, \text{where } c\sim\mathcal{N}(0,\sigma^2)\;,
\end{equation}
and $x$ is the distance between two ends, and $a$, $b$ are constants. We consider using Internet as the backhaul connecting primary users, secondary users, and the spectrum manager(s). In \cite{goonatilake2012}, the parameters are found to be $a=0.022$, $b=4.862$, and $\sigma^2=0.907$, where $x$ and $\delta(x)$ are in miles and milliseconds, respectively. 

The second type of delay is the response time of the spectrum manager between receiving the signaling message of PU and a spectrum scheduling command is sent out to target secondary users, which is further modeled by cyber model in the next section. 

The third type of delay is the evacuation time of secondary link, as illustrated by an exemplary process in Fig.~\ref{fig:delay:su}. Suppose the secondary link are operated in a time division mode, which is the alternating downlink and uplink transmissions on channel $b$. The downlink is transmission from secondary base-station (SBS) to secondary user equipment (SUE), and vice versa the uplink. The spectrum scheduling command is always sent to SBS, which is directly connected to the backhaul network. The command however can arrive at an arbitrary time of the normal operating of the secondary link. If a new frame of downlink and uplink already began, (indicated as close of a scheduling window in Fig.~\ref{fig:delay:su}), the SBS will wait until the next downlink and uplink frame. Since the arrival of command and the secondary link are independent, we assume a uniformly distributed random delay $e$ is incurred until a handover process starts. The handover process can be viewed as a conversation between SBS and SUE that include a DL and an UL packets. In the DL packet, the SBS tells SUE to handover from channel $b$ to the next free channel $k$, and ST returns an acknowledge in consequent UL packet. After the UL packet, the SBS and ST stop transmission on channel $b$, which marks a event 2. 
As a result, the handover delay is modeled as:
\begin{equation}\label{eq:handover}
	t_H = e + f \;,
\end{equation}
where $f$ is the fixed overhead, and $e\sim\mathcal U(0,l_f)$ is a uniform random variable between two frames, where $l_f$ is the frame length. 

Since the secondary link interferes the TV receiver, it is only when the secondary link evacuated channel $b$ (event 2), the TV receiver could start to buffer and decode the TV signal on channel $b$. The real-time constraints of the physical model is that time between a zapping (event 1) and according evacuation (event 2), which is the evacuation delay, $\Delta_{e}$,  should not exceed an threshold $Pr(\Delta_{e}\leq\Delta_{max}) = O_{max}$, where $O_{max}$ is a prescribed protection probability. For example, the digital TV set usually takes 1-2 seconds to buffer and decode the TV signal before the first frame of image is displayed on the screen \cite{Lee_Reducing_2010}, which is called zapping time. The evacuation delay increases the zapping time, hence degrades the quality of experience (QoE) of the TV user. Assuming a 10-20\% increase of zapping time is acceptable to the TV user, then the evacuation time should be lower than 200ms, $\Delta_{max}\leq200ms$ \cite{Rempe2017,zhao17}.

\subsection{Cyber Model}\label{sec:cps:cyber}

\begin{figure} [h!]
	\centering 
	\subfloat[]{
	\includegraphics[height=1.6in]{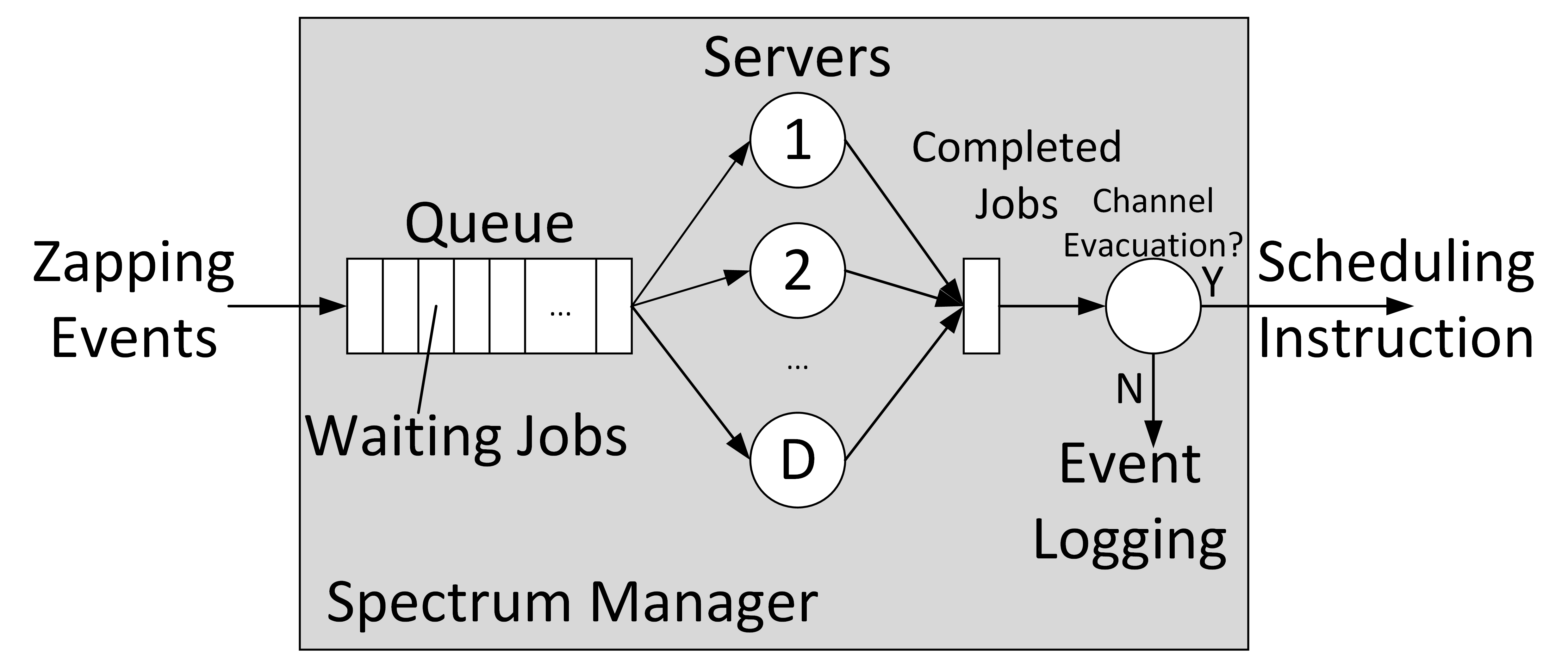}
	\label{fig:delay:queue}
	}
	\subfloat[]{
		\includegraphics[height=1.3in]{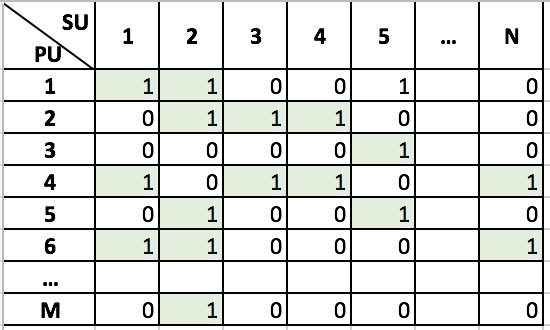}
		\label{fig:delay:db}
	}
	\caption{Cyber Model at Spectrum Manager: (a) Queueing Model, (b) An Exemplary Interference Relation Database.}\label{fig:cyber}
\end{figure}

The processing of arriving signaling message at the spectrum manager is modeled as a queueing model $M/G/C$, as shown in Fig.~\ref{fig:delay:queue}. The $M/G/C$ queueing model is explained as follows: The arrival process of signaling message triggered by zapping events is assume to follow a Poisson process, denoted by $M$. The arrival rate of zapping events is proportional to the number of TV receivers registered at the spectrum manager, and the activity levels of these TV receivers. The arriving signaling messages is put in a queue, and be processed based on a first-arrive-first-serve rule. There are $C$ servers within the spectrum manager, e.g. multi-core, or cluster, processing these zapping messages. When a job is complete, there are two possible outcomes: either some secondary users need to evacuate the channel, or no secondary user need to evacuate the channel. The latter happens when there already many TV users watching that TV channel the TV user is demanding, or there are no secondary user operating on that TV channel in the protection zone of the TV user. For the former outcome, a spectrum scheduling command will be sent to the affected secondary user. For the latter outcome, the spectrum manager only need to log or update its database. Based on the queueing model in Fig.~\ref{fig:delay:queue}, the processing time of an arriving zapping events is determined by how many active TV receivers is registered at the spectrum manager. The more TV receivers a spectrum manager handles, the longer the waiting time in the queue, and the longer the job processing time. 

The job processing time is determined by the numbers of primary users and secondary users the spectrum manager handles due to the size effect of a relational database. The relational database within the spectrum manager, as shown in Fig.~\ref{fig:delay:db}, records the interference relationship between secondary users and primary users. The interference relationship is mainly determined by the physical distance between the secondary and primary users, as well as the radio environments. For each TV receiver, there is a prescribed guard zone, which is a disk centered at the TV receiver, with radius, $r_p$. Any secondary user within the guard zone of the TV receiver is not allowed to transmit on the channel being watched by that TV receiver. Therefore, based on the geolocations of PUs and SUs that are known to the spectrum manager, a relational database can be established. In the exemplary interference database shown in Fig.~\ref{fig:delay:db}, there are N SUs and M PUs, and SUs 1, 2, 5, are in the guard zone of PU 1, and so on. Notice that SUs and PUs are dynamic as they can be added to or removed from the system.

When a spectrum management job is processed, the spectrum manager will first search the interference database for all the SUs that within the guard zone of the requesting PU (of which the identity is included in the signaling message). Then the spectrum manager checks the channels occupied by these selected SUs to see if the  channel requested by PU is occupied or not. The time complexity of this process is $\mathcal{O}(M+N)$. If there are SUs need to be evacuated, the spectrum manager need to find out a list of free channels for them to continue their operations. The spectrum manager will search the interference database to find out all the PUs that is interfered by an secondary user, and find out the channels that are not used by those PUs. The time complexity of finding available channel for each interrupted SU link is $\mathcal{O}(M)$. The spatial distribution of SUs is modeled as Poisson Point Process, with density $\lambda_{s}$ \cite{zhao17}. The number of SUs in the guard zone of each TV receiver, $n$, follows a Poisson distribution, with Probability Mass Function (PMF):
\begin{equation}\label{eq:poisson:su}
Pr(n=k)=\frac{(\lambda_{s}\pi r_p^2)^{k}\exp(-\lambda_{s}\pi r_p^2)}{k!}\;.
\end{equation}
Similarly, spatial distribution of TV receives is also modeled as a Poisson Point Process with density $\lambda_{p}$, and the number of PUs in the interference range of a SBS, $m$, also follows a Poisson distribution, with Probability Mass Function:
\begin{equation}\label{eq:poisson:pu}
Pr(m=k)=\frac{(\lambda_{p}\pi r_p^2)^{k}\exp(-\lambda_{p}\pi r_p^2)}{k!}\;,
\end{equation}

The service time of a spectrum management job that involves event 2 can be expressed as:
\begin{equation}
t_p = g(M+N) + gMn + hmn + l\;,
\end{equation}
where $g$ is the unit time consumption for searching an item in the database, $h$ is the unit time of looking up channels occupied by a PU, and $l$ is a random delay caused by internal uncertainty of the computing system, such as scheduling delay of Operating System. In \cite{matalqa2016effect}, the average response time of a query in disk-based database for dataset size of 1,200, 4,800, and 19,000 is reported as 6, 26, and 220 milliseconds, respectively. For telecommunications, fast response is achieved by in-memory database \cite{Jose2013,HZhang15}. For example, IBM solidDB \cite{lindstrom2013ibm}, can reduce the response time from 375 milliseconds in disk-based database to 50 milliseconds, and achieve an average of 1.2 million transactions per second for a dataset of 1 million records. In \cite{Jose2013}, SolidDB is reported to achieve 140,000 transactions per seconds on Flash-backed DRAM storage device.

\section{Scale-based Evacuation Delay Model}\label{sec:model}
The evacuation delay is the sum of network latency in \eqref{eq:network}, spectrum handover delay in \eqref{eq:handover}, and response time of spectrum manager. Since the total time a job spent in the queueing system in Section \ref{sec:cps:cyber} does not have a closed-form, an approximation is found first, then the total evacuation delay is modeled based on the approximated response time.
\subsection{Approximation of Spectrum Manager Response Time}
From the cyber model in Section \ref{sec:cps:cyber}, the service time of each arriving zapping event is proportional to the number of SUs need to be evacuated. The upper bound of evacuated SUs is the number of SUs, $n$, in the guard zone of that TV receiver. For tractability, the service time $t_p$ is approximated to follow an exponential distribution \cite{KIMURA1995,PSOUNIS2005} with a mean of $\tau E(n)$, where $\tau$ is a constant, and $E(n) = \lambda_{s}\pi r_p^2$.

The spectrum manager response time $t_D$ is represented as the total delay of the $M/M/C$ queueing system. The Laplace transform of the PDF of the response time is \cite{giambene2005queuing}
\begin{equation}\label{eq:laplace:td}
T_D(s) = (1-P_C)\frac{\mu}{\mu+s} + P_C\frac{(\mu C-\lambda)\mu}{(\mu C-\lambda+s)(\mu+s)}\;,
\end{equation}
where $P_C$ denotes the Erlang-C formula \cite{giambene2005queuing}:
\begin{equation}\label{eq:erlangc}
	P_C = \frac{\frac{C\rho^C}{C!(S-\rho)}}{\sum_{i=0}^{C-1}\frac{\rho^i}{i!}+ \frac{C\rho^C}{C!(C-\rho)}}\;,
\end{equation}
$\lambda$ is the arrival rate, and $\mu$ is the service rate, and the intensity of the arrival process $\rho = \lambda/\mu$. The service rate $\mu=1/{\left(\tau\lambda_{s}\pi r_p^2\right)}$, and the arrival rate can be found by \cite{zhao17}
\begin{equation}
\lambda = \frac{M \phi(t)}{E(B)}\;,
\end{equation}
where $M$ is the number of registered TV sets at the spectrum manager, $E(B)$ is the average channel holding time of TV users \cite{Cha2008}, $\phi(t)$ is the Household Using Television (HUT), which is the ratio of active TV receiver to all TV receivers \cite{zhao17}.

From this approximation, the distribution of response time of the Spectrum Manager is determined by the number of registered TV sets at the spectrum manager $M$, the density of SUs $\lambda_{s}$, size of guard zone $r_{p}$, number of servers at the spectrum manager, average channel holding time $E(B)$, and HUT which represents the activity level of TV users.

\subsection{Distribution of Evacuation Delay}\label{sec:model:total}
The overall evacuation delay of a secondary user, $t_E$, can be modeled as the sum of network latency, $t_N$, response time of spectrum manager, $t_D$, and spectrum handover overhead, $t_H$, as follows:
\begin{equation}\label{eq:sum}
t_E = t_N + t_D + t_H\;.
\end{equation}
Since $t_N$, $t_D$, and $t_H$ are all random variables, the Laplace transform of the PDF of $t_E$ can be expressed as the product of Laplace transform of PDFs of $t_N$, $t_D$, and $t_H$:
\begin{align}
T_E(s) = &T_N(s) T_H(s) T_D(s) \\
= &\exp\left(s(ax+b)+\frac{s^2}{2}\sigma^2\right) + \frac{e^{-sf}-e^{-s(f+l_f)}}{sl_f}
 + (1-P_C)\frac{\mu}{\mu+s} + P_C\frac{(\mu C-\lambda)\mu}{(\mu C-\lambda+s)(\mu+s)} \;.
\end{align}
For fully centralized solutions, the $x$ is considered to be the distance between locations of spectrum manager (e.g. a data center) and the furtherest cities.  $a$ and $b$ are constants in the network latency model, $\sigma^2$ is the variance of random component $c$ in the network latency model. $f$ is the fixed time consumption of spectrum handover, $l_f$ is the frame length in the air interface protocol. For regional spectrum manager located in the city where TV users under its charge, $x=0$. 


The scale of of PUs and SUs handled by a single spectrum manager also determines the distribution of physical distances between the PUs and spectrum manager. Since we consider the guard zone of PU is small (within hundreds of meters), for regional spectrum manager, the distance between secondary link and spectrum manager is assumed identical to the distance between PU and the spectrum manager. However, if a fully distributed solution is employed, the distance between SBS and spectrum manager is 0, and $N=1$.

\section{Numerical Evaluation}\label{sec:eval}
We evaluate 4 solutions based on assumed and collected parameters of aforementioned network latency model, frame length, and evacuation model. The four solutions are: 
\begin{itemize}
	\item Fully distributed solution where each secondary access point handles the spectrum management functionality for itself. The number of processors of spectrum manager is set to 1.
	\item Regional Centralization, where a spectrum manager taking care of all the TV receivers in the city it located in. The number of processors in the spectrum manager is assumed to be 32.
	\item National Centralization, where a single spectrum manager located in Lincoln, Nebraska (approximately the geometric center of continental U.S.) taking care of all the TV receivers in the U.S. The number of processors in the spectrum manager is assumed to be 100,000, since it might be a data center.
	\item Semi-National Centralization, where 50 individual spectrum managers co-located in Lincoln, Nebraska, taking care of all the TV receivers in the U.S. Each spectrum managers only take care of at most 1 million TV receivers. The number of processors is assumed to be 10,000 for each individual spectrum manager.
\end{itemize} 
Based on the statistics of network delay from Internet Service Providers \cite{windstream17,verizon17,att17}, we consider the average round trip latency from coastal cities to Lincoln is 25 ms, and the average round trip latency within the same city as 3ms. The variance of network latency is $\sigma^2=0.907$. The database response time is considered to be 6--50ms when the records in the database is equal to or less than 1 million. The Over-The-Air TV ownership rate is set to 14\%. A city of 8 million population (1.12 million Over-The-Air TV receivers) is considered for the regional spectrum manager case. National wide, the number of Over-The-Air TV receivers is then 44.8 million. The TV usage level is considered to vary from 5\% to 60\% from the middle night to prime time in a typical day. We assume that on average there are 10 to 200 secondary users in the guard zone (radius of 130 meters) of each primary user. The frame length of the communication protocol is assumed to be 20 ms, which is the maximum frame length of WiFi network. As a result, in this evaluation, the spectrum handover overhead of secondary link follows uniform distribution, $t_H\sim\mathcal{U}(20ms, 40ms)$.

\begin{table*}[!t]
	\renewcommand{\arraystretch}{1}
	\vspace{-0.1in}
	\caption{Summary of Average Channel Evacuation Time of Four Levels of Centralization} 
	\label{tab:summary}
	\centering
	\begin{tabular}{|c||c|c|c|c|c|}
		\hline
		Centralization & No. of Processor & No. of TV Receivers & Network Latency & SM response time & Evacuation Time \\ \hline\hline		
		Fully Distributed & 1 & $\leq130$ & 2 ms & 10 ms & 35 to 55 ms   \\ \hline
		Regional Centralization & 32 &  1.12 million &  5 ms & 120 ms & 155 ms  \\ \hline
		National Centralization. & 100,000 & 45 million & 25 ms & 2000 ms & 2060 ms   \\ \hline
		Semi-National Centralization & 10,000 & 1 million & 25 ms &  120 ms & 180 ms   \\ \hline
	\end{tabular}
\end{table*}

\subsection{Fully Distribution}
In the fully distributed scenario, the average network latency is 3ms. In this case, the number of registered TV receivers at each spectrum manager (in this case an individual secondary access point) is less than or equal to 130. This means that the relational database at the spectrum manager has at most a few hundred records. The response time of this tiny relational database will be fastest, 10 ms. Based on these conditions, we can find the average evacuation time for the fully distributed spectrum manager is 32 to 55 ms. 

\subsection{Regional Centralization}
In the regional centralization scenario, we assume that one spectrum manager located inside the city handles all the TV receivers in a city. In this setting, the number of TV receivers is assumed to be 1.12 million. The number of secondary users in the guard zone of TV receiver is 10-200. The spectrum manager is assumed to be a small workstation, which has 32 processors. We assume that the spectrum manager use in memory database which has the fastest response among all the database solutions. The response time of a single query is 6ms, and the response time average job will handle 200 records, with the probability of 0.1 for secondary users to evacuate the channel. Therefore, each job at the spectrum manager that will issue evacuation instruction will take about $120=6\times200\times0.1$ ms. The average evacuation time $t_E$ is 155 ms in regional centralized spectrum manager. 

\subsection{National Centralization}
For the national centralization, the average network delay from the spectrum manager to coastal cities is 25ms. The number of TV receivers in the interference relation database is 45 million. The number of secondary users in the guard zone of TV receiver is 10-200. With database of 45 million records, the response time of individual query will be longer, 100 ms. The job at the spectrum manager that will issue channel evacuation instruction will take $2000=100\times200\times0.1$ ms. 

\subsection{Semi-National Centralization}
Finally, we look at the semi-national centralization solution. The network latency is the same as national centralization scenario, which is 25 ms round trip latency.  The difference is that the size of the interference relation database at each individual spectrum manager is limited to 1 million records. Therefore,  an individual database query will be responded in shorter time, which is the same as the regional centralization case, $120$ ms. The average evacuation time is then 180 ms, which is longer than the regional centralization scenario due to additional network latency. 

The average evacuation time in the four centralization scenarios is summarized in Table \ref{tab:summary}.

\section{Conclusion}
Through a preliminary analysis of the four architecture of spectrum manager in TV set-assisted spectrum sharing, we find that except the fully centralization, the solutions of fully distributed, regionally centralized, and semi-nationally centralized spectrum manager can all meet the real-time requirement of evacuating the TV channels for TV user within 200-300 ms. From the operational perspective, semi-national centralized solution provides best privacy protection of TV users, as well as all the benefit of cloud computing in terms of maintenance, management, energy consumption, and reduced cost of secondary access point.

Future research may include verification of the assumed response time, distribution of network latency, and evaluate the analytical model. The preliminary analysis presented in this paper only include the average latency. Future research should include distributions of the evacuation time, and evaluate if the distribution meet the regulatory requirement. Finally, we also should consider the failure rate of Internet. In this case, centralized solution might be in disadvantage compared to fully distributed solution. 

\section*{Acknowledgment}
This work is supported by National Science Foundation under grant number NSF CNS-1247941 and CNS-1247914. The author would also thank Dr. Justin M. Bradley and Dr. Mehmet C. Vuran for their invaluable guidance on this technical report.

\bibliographystyle{IEEEtran}
\bibliography{CogTV,CPS}

\begin{thebibliography}{10}
\providecommand{\url}[1]{#1}
\csname url@samestyle\endcsname
\providecommand{\newblock}{\relax}
\providecommand{\bibinfo}[2]{#2}
\providecommand{\BIBentrySTDinterwordspacing}{\spaceskip=0pt\relax}
\providecommand{\BIBentryALTinterwordstretchfactor}{4}
\providecommand{\BIBentryALTinterwordspacing}{\spaceskip=\fontdimen2\font plus
\BIBentryALTinterwordstretchfactor\fontdimen3\font minus
  \fontdimen4\font\relax}
\providecommand{\BIBforeignlanguage}[2]{{%
\expandafter\ifx\csname l@#1\endcsname\relax
\typeout{** WARNING: IEEEtran.bst: No hyphenation pattern has been}%
\typeout{** loaded for the language `#1'. Using the pattern for}%
\typeout{** the default language instead.}%
\else
\language=\csname l@#1\endcsname
\fi
#2}}
\providecommand{\BIBdecl}{\relax}
\BIBdecl

\bibitem{Akyildiz2006}
I.~F. Akyildiz, W.-Y. Lee, M.~C. Vuran, and S.~Mohanty, ``{NeXt}
  generation/dynamic spectrum access/cognitive radio wireless networks: A
  survey,'' \emph{Comput. Netw.}, vol.~50, no.~13, pp. 2127--2159, Sep 2006.

\bibitem{2ndOrder}
``Second memorandum opinion and order,'' Federal Communications Commission, Sep
  2010.

\bibitem{5457914}
K.~Harrison, S.~Mishra, and A.~Sahai, ``How much white-space capacity is
  there?'' in \emph{IEEE DySPAN}, Apr. 2010, pp. 1--10.

\bibitem{Hessar2014}
F.~Hessar and S.~Roy, ``Capacity considerations for secondary networks in {TV}
  white space,'' \emph{IEEE Trans. on Mobile Computing}, vol.~14, no.~9, pp.
  1780--1793, Sep. 2015.

\bibitem{zhzhao_techreport}
Z.~Zhao and M.~C. Vuran, ``Population density statistics and {OTA TV} channel
  availability via {Cog-TV} in major united states cities,'' University of
  Nebraska-Lincoln, Tech. Rep. TR-UNL-CSE-2014-0006, Dec 2014,
  \url{http://www.unl.edu}.

\bibitem{Fay2016}
L.~Fay, L.~Michael, D.~Gómez-Barquero, N.~Ammar, and M.~W. Caldwell, ``An
  overview of the atsc 3.0 physical layer specification,'' \emph{IEEE
  Transactions on Broadcasting}, vol.~62, no.~1, pp. 159--171, March 2016.

\bibitem{Zhao14Globecom}
Z.~Zhao, M.~C. Vuran, D.~Batur, and E.~Ekici, ``Ratings for spectrum: Impacts
  of {TV} viewership on {TV} whitespace,'' in \emph{IEEE GLOBECOM 2014}, Dec
  2014, pp. 941--947.

\bibitem{Rempe2017}
D.~Rempe, M.~Snyder, A.~Pracht, A.~Schwarz, T.~Nguyen, M.~Vostrez, Z.~Zhao, and
  M.~C. Vuran, ``A cognitive radio tv prototype for effective tv spectrum
  sharing,'' in \emph{IEEE DySPAN}, Mar. 2017, pp. 117--118.

\bibitem{zhao2017}
Z.~Zhao, M.~C. Vuran, D.~Batur, and E.~Ekici, ``Shades of white: Impacts of
  population dynamics and {TV} viewership on available {TV} spectrum,''
  \emph{IEEE Trans. on Vehicular Technology}, May 2018, submitted to.

\bibitem{LiQiao10}
Q.~Li, ``Scheduling in cyber physical systems,'' Ph.D. dissertation, Carnegie
  Mellon University, Aug. 2012.

\bibitem{goonatilake2012}
R.~Goonatilake and R.~A. Bachnak, ``Modeling latency in a network
  distribution,'' \emph{Network and Communication Technologies}, vol.~1, no.~2,
  p.~1, 2012.

\bibitem{Lee_Reducing_2010}
C.~Y. Lee, C.~K. Hong, and K.~Y. Lee, ``Reducing channel zapping time in iptv
  based on user's channel selection behaviors,'' \emph{IEEE Transactions on
  Broadcasting}, vol.~56, no.~3, pp. 321--330, Sept 2010.

\bibitem{zhao17}
Z.~Zhao, M.~Vuran, D.~Batur, and E.~Ekici., ``Shades of white: Impacts of
  population dynamics and tv viewership on available tv spectrum,''
  \emph{Submitted to IEEE Transactions on Wireless Communication}, 2017.

\bibitem{matalqa2016effect}
S.~Matalqa and S.~Mustafa, ``The effect of horizontal database table
  partitioning on query performance.'' \emph{Int. Arab J. Inf. Technol.},
  vol.~13, no.~1A, pp. 184--189, 2016.

\bibitem{Jose2013}
\BIBentryALTinterwordspacing
J.~Jose, M.~Banikazemi, W.~Belluomini, C.~Murthy, and D.~K. Panda, ``Metadata
  persistence using storage class memory: Experiences with flash-backed dram,''
  in \emph{Proceedings of the 1st Workshop on Interactions of NVM/FLASH with
  Operating Systems and Workloads}, ser. INFLOW '13.\hskip 1em plus 0.5em minus
  0.4em\relax New York, NY, USA: ACM, 2013, pp. 3:1--3:7. [Online]. Available:
  \url{http://doi.acm.org/10.1145/2527792.2527800}
\BIBentrySTDinterwordspacing

\bibitem{HZhang15}
H.~Zhang, G.~Chen, B.~C. Ooi, K.~L. Tan, and M.~Zhang, ``In-memory big data
  management and processing: A survey,'' \emph{IEEE Transactions on Knowledge
  and Data Engineering}, vol.~27, no.~7, pp. 1920--1948, July 2015.

\bibitem{lindstrom2013ibm}
J.~Lindstr{\"o}m, V.~Raatikka, J.~Ruuth, P.~Soini, and K.~Vakkila, ``Ibm
  soliddb: In-memory database optimized for extreme speed and availability.''
  \emph{IEEE Data Eng. Bull.}, vol.~36, no.~2, pp. 14--20, 2013.

\bibitem{KIMURA1995}
\BIBentryALTinterwordspacing
T.~Kimura, ``Approximations for the delay probability in the m/g/s queue,''
  \emph{Mathematical and Computer Modelling}, vol.~22, no.~10, pp. 157 -- 165,
  1995. [Online]. Available:
  \url{http://www.sciencedirect.com/science/article/pii/0895717795001925}
\BIBentrySTDinterwordspacing

\bibitem{PSOUNIS2005}
\BIBentryALTinterwordspacing
K.~Psounis, P.~Molinero-Fernández, B.~Prabhakar, and F.~Papadopoulos,
  ``Systems with multiple servers under heavy-tailed workloads,''
  \emph{Performance Evaluation}, vol.~62, no.~1, pp. 456 -- 474, 2005,
  performance 2005. [Online]. Available:
  \url{http://www.sciencedirect.com/science/article/pii/S0166531605001094}
\BIBentrySTDinterwordspacing

\bibitem{giambene2005queuing}
G.~Giambene, \emph{Queuing theory and telecommunications}.\hskip 1em plus 0.5em
  minus 0.4em\relax Springer, 2005.

\bibitem{Cha2008}
M.~Cha, P.~Rodriguez, J.~Crowcroft, S.~Moon, and X.~Amatriain, ``Watching
  television over an ip network,'' in \emph{Proc. of the 8th ACM SIGCOMM
  Conference on Internet Measurement}, ser. IMC '08.\hskip 1em plus 0.5em minus
  0.4em\relax New York, NY, USA: ACM, 2008, pp. 71--84.

\bibitem{windstream17}
\BIBentryALTinterwordspacing
``Windstream's real-time network latency,'' Windstream Wholesale, retrived by
  Nov. 20th, 2017. [Online]. Available:
  \url{http://www.windstreamwholesale.com/network-latency-tool/}
\BIBentrySTDinterwordspacing

\bibitem{verizon17}
\BIBentryALTinterwordspacing
``Ip latency statistics,'' Verizon. [Online]. Available:
  \url{http://www.verizonenterprise.com/about/network/latency/#latency}
\BIBentrySTDinterwordspacing

\bibitem{att17}
\BIBentryALTinterwordspacing
``U.s. network latendy,'' AT\&T. [Online]. Available:
  \url{http://ipnetwork.bgtmo.ip.att.net/pws/network_delay.html}
\BIBentrySTDinterwordspacing

\end{thebibliography}

\end{document}